\documentclass[submission, Proceedings]{SciPost}

\hypersetup{
    colorlinks,
    linkcolor={red!50!black},
    citecolor={blue!50!black},
    urlcolor={blue!80!black}
}

\usepackage[bitstream-charter]{mathdesign}
\usepackage{cleveref}
\usepackage{verbatim}
\usepackage{wrapfig}

\DeclareSymbolFont{usualmathcal}{OMS}{cmsy}{m}{n}
\DeclareSymbolFontAlphabet{\mathcal}{usualmathcal}

\newcommand{\unit}[1]
{
\;\mathrm{#1}
}

\begin{document}

\begin{center}{\Large \textbf{
Measurements of $W$ and $Z/\gamma^*$ cross sections and their ratios in $pp$ collisions at STAR\\
}}\end{center}

\begin{center}
Jae D. Nam\textsuperscript{1,*} for the STAR collaboration
\end{center}

\begin{center}
{\bf 1} Temple University, Philadelphia, U.S.
\\
* jae.nam@temple.edu
\end{center}

\begin{center}
\today
\end{center}


\definecolor{palegray}{gray}{0.95}
\begin{center}
\colorbox{palegray}{
  \begin{tabular}{rr}
  \begin{minipage}{0.1\textwidth}
    \includegraphics[width=22mm]{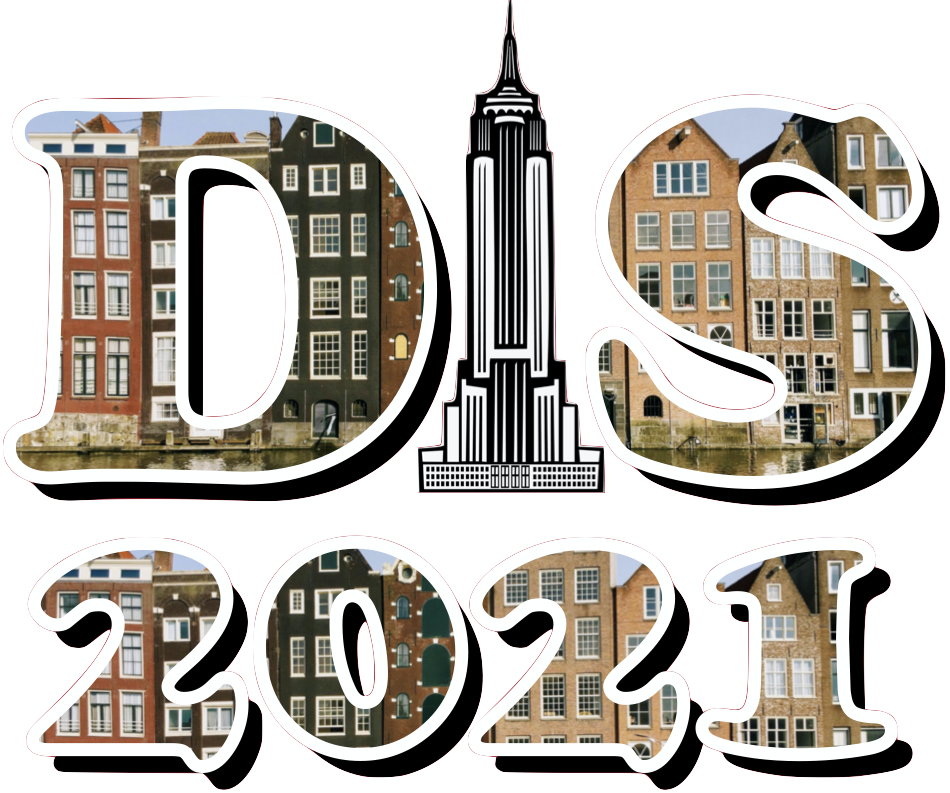}
  \end{minipage}
  &
  \begin{minipage}{0.75\textwidth}
    \begin{center}
    {\it Proceedings for the XXVIII International Workshop\\ on Deep-Inelastic Scattering and
Related Subjects,}\\
    {\it Stony Brook University, New York, USA, 12-16 April 2021} \\
    \doi{10.21468/SciPostPhysProc.?}\\
    \end{center}
  \end{minipage}
\end{tabular}
}
\end{center}

\section*{Abstract}
{\bf
While the unpolarized parton distribution functions of the valence quarks ($d$ and $u$) are well determined from DIS experiments, the sea quark counterparts, $\bar{d}$ and $\bar{u}$, are much less constrained, in particular, near the valence region.
Measurements of $W^+/W^-$ production ratio in $pp$ collider experiments, such as the STAR experiment at RHIC, can probe the $\bar{d}/\bar{u}$ ratio at a large $Q^2$ set by the $W$ mass.
These proceedings will discuss recent results of $W$ and $Z$ cross-section measurements using the STAR $pp$ collision data at a center-of-mass energy of $\sqrt{s} = 500/510\;\mathrm{GeV}$ collected in 2011, 2012, 2013 and 2017.
}



\section{Introduction}
\label{sec:intro}

Over the past decades, experiments at high-energy particle colliders have proven successful in broadening our understanding of the nucleon structure and constraining the parton distribution functions (PDFs) more precisely than before.
Several global analyses, such as CT14~\cite{Gao:2013xoa}, MMHT14~\cite{Szydlowski:2008by}, and BS15~\cite{Bourrely:2015kla}, have been conducted in order to extract PDFs from experimental observations.
Nevertheless, there remain open questions that require further experimental studies.
One example is the unpolarized sea quark distribution ratio $\bar{d}/\bar{u}$, which has been predominantly extracted from Drell-Yan measurements, such as those at E866~\cite{NuSea:2001idv} and the SeaQuest~\cite{SeaQuest:2021zxb} experiments.
While measurements from the two experiments are in good agreement in the region of the partonic momentum fraction $x < 0.2$, a discrepancy is observed for $x > 0.25$.
This motivates complementary measurements using different physics processes.

A strong candidate for such a process is the $W$ production in $pp$ collisions, where the $W^+/W^-$ cross-section ratio is sensitive to the sea quark distribution ratio.
The production and decay channel of a $W^+$ ($W^-$) boson is shown in equation~\ref{e1}.

\begin{equation}
\label{e1}
u + \bar{d} \rightarrow W^+ \rightarrow e^+ + \nu \qquad (d + \bar{u} \rightarrow W^- \rightarrow e^- + \bar{\nu})
\end{equation}

\noindent
At the leading order, the $W$ cross-section ratio, $\sigma_{W+}/\sigma_{W-}$, is directly proportional to the sea quark PDFs, as shown in equation~\ref{e2}~\cite{Soffer:2014upa}, and can be used to probe the sea quark distributions at a large $Q^2 \sim M_W^2$.

\begin{equation}
\label{e2}
\frac{\sigma_{W+}}{\sigma_{W-}} \sim \frac{\bar{d}(x_2) u(x_1) + \bar{d}(x_1) u(x_2)}{\bar{u}(x_2) d(x_1) + \bar{u}(x_1) d(x_2)}
\end{equation}

\noindent
Measurements of $W$ production at STAR are also complementary to the LHC $W$ and $Z$ production measurements by providing cross-section measurements with lower $\sqrt{s}$ and larger $x$.

\section{Experiment}
\label{sec:experiment}

A detailed description of the STAR detector can be found in~\cite{STAR:2002eio}. A brief outline of the components that are most relevant for this analysis is given below.

The time projection chamber (TPC)~\cite{Anderson:2003ur} is a drift chamber used for particle tracking,
and the barrel electromagnetic calorimeter (BEMC)~\cite{STAR:2002ymp} and endcap electromagnetic calorimeter (EEMC)~\cite{STAR:2002eml}  are used to measure electrons from weak boson decays and for triggering.
In the mid-rapidity region ($|\eta| \leq 1$) STAR probes the $x$ range of approximately $0.1$ to $0.3$.
The EEMC allows the $W$ cross-section ratio to be measured in the more forward direction, which extends the $x$ reach to roughly $0.06 < x < 0.4$.

The first measurements of $W$ and $Z$ cross sections at STAR were based on a data set taken in 2009, corresponding to an integrated luminosity of $\mathcal{L} \sim 13.2 \unit{pb^{-1}}$~\cite{STAR:2011aa}, at a center-of-mass energy $\sqrt{s} = 500 \unit{GeV}$.

Additional data sets were collected in 2011 at $\sqrt{s} = 500 \unit{GeV}$, and 2012 and 2013 at $510 \unit{GeV}$ with a combined luminosity of $\mathcal{L} \sim 350 \unit{pb^{-1}}$.
The results based on these data sets have been published in~\cite{STAR:2020vuq}.
Finally, a preliminary release has been granted based on the data set collected in 2017 at $\sqrt{s} = 510 \unit{GeV}$ with $\mathcal{L} \sim 350 \unit{pb^{-1}}$.

\section{Results}
\label{sec:results}

$W$ candidates are selected by requiring a large imbalance in the transverse momentum ($p_\mathrm{T}$) deposit in the detector, arising from final-state neutrinos that escape the STAR detector system. Furthermore, the decay-electron tracks are tagged by requiring a highly-concentrated energy deposition in either BEMC or EEMC.

\begin{figure}[h]
\centering
\includegraphics[width=0.35\textwidth]{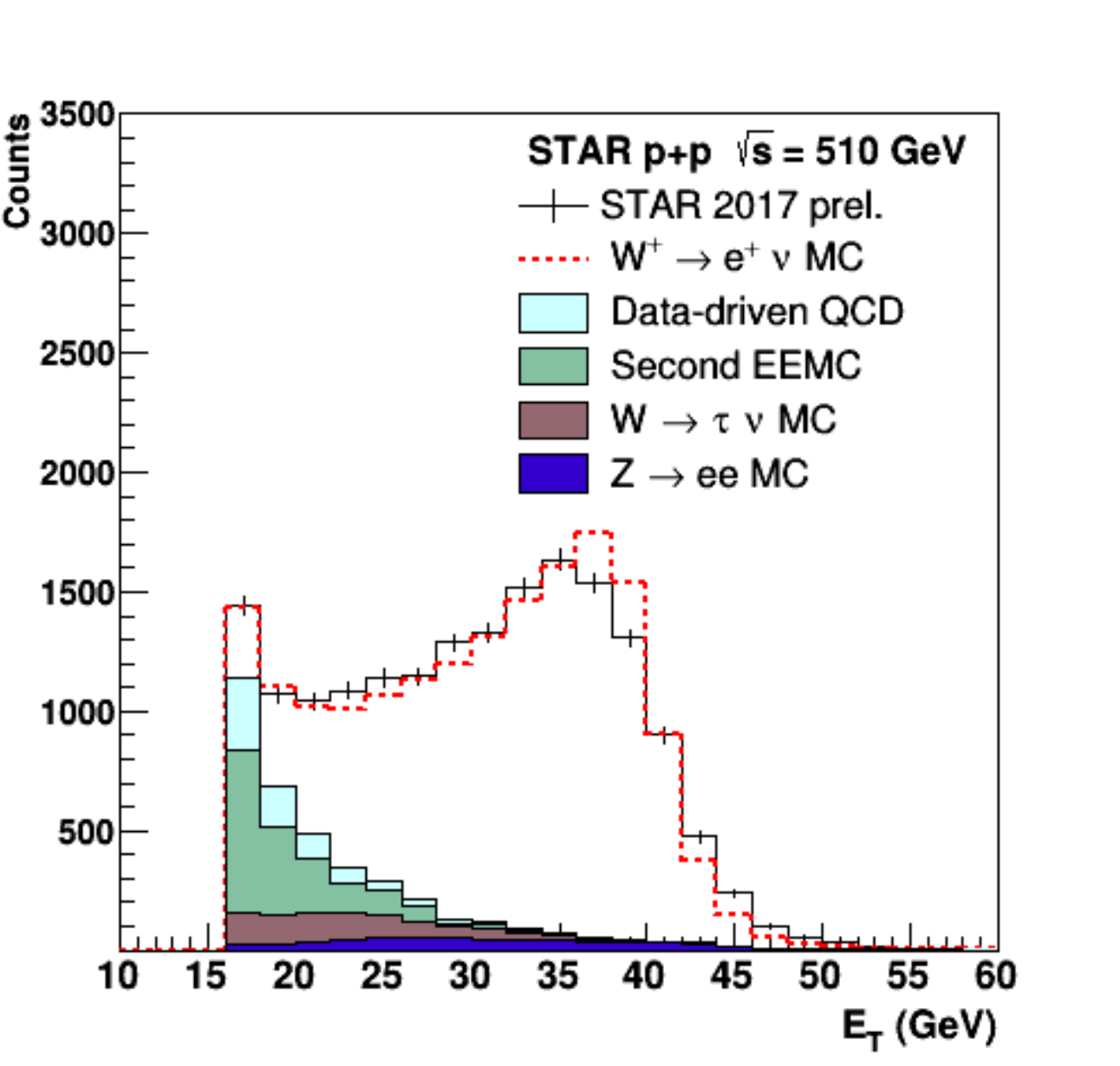}
\includegraphics[width=0.35\textwidth]{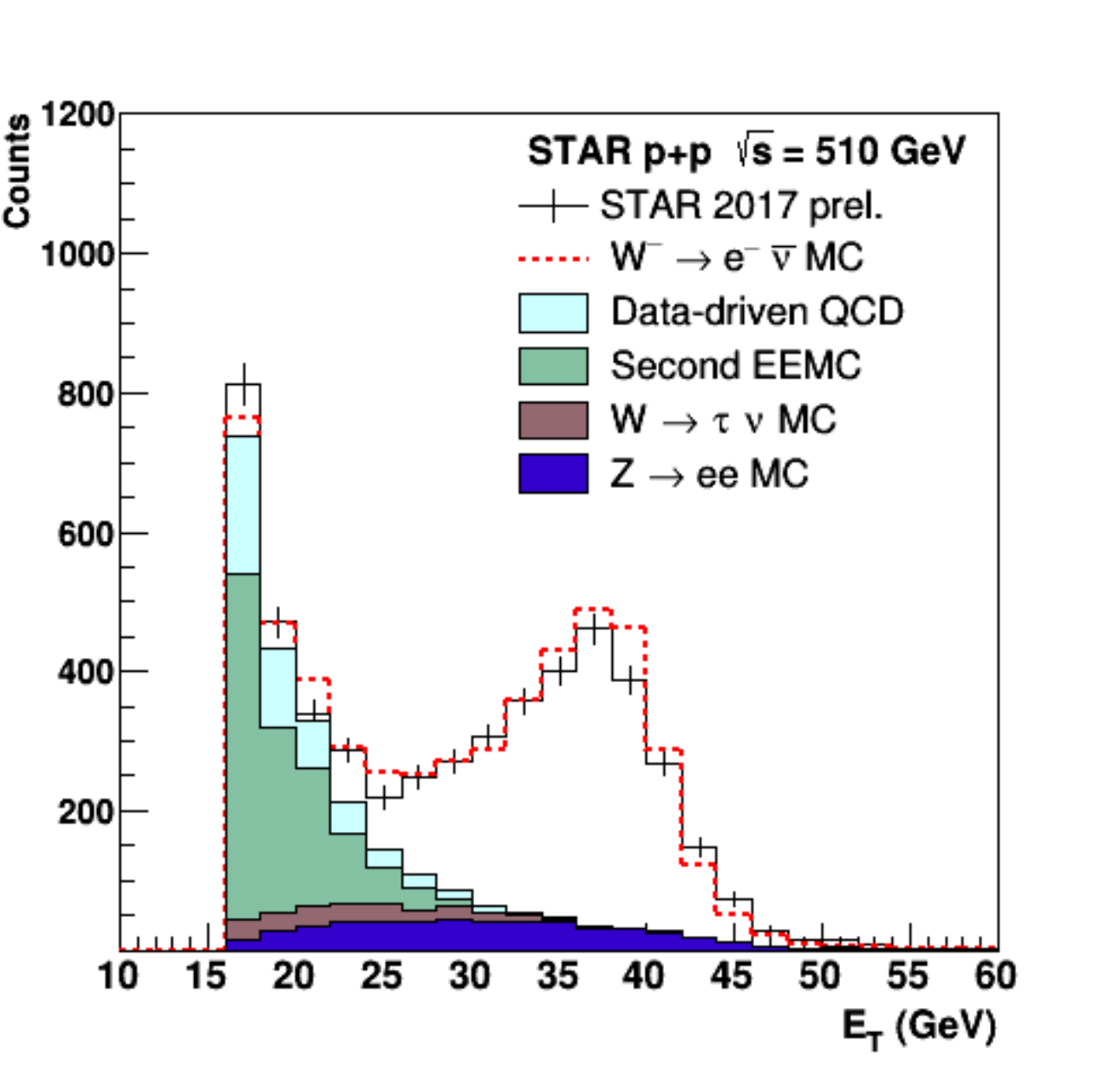}
\caption{
$E_\mathrm{T}$ distributions for electroweak $e^{+}$ (left) and $e^{-}$ candidates (right).
}
\label{fig:ETeW}
\end{figure}

A comparison of the electroweak (EW) $e^\pm$ candidates from the STAR 2017 data set and a Monte-Carlo (MC) simulation (based on Pythia 6.4.28~\cite{Sjostrand:2006za} and GEANT~\cite{GEANT4:2002zbu}) is given as a distribution of transverse energy ($E_\mathrm{T}$) in Fig.~\ref{fig:ETeW}.
The MC distribution is a combination of the EW electron signal and various background contributions;
$W \rightarrow e\nu$ MC represents the $W$-decay electron signal from the MC simulation;
the QCD contribution that survives the selection cuts was estimated from the $E_\mathrm{T}$ distribution that fails the $p_\mathrm{T}$ imbalance cut, and is denoted in the figure as data-driven QCD;
the second EEMC background is an estimate of the non-$W$ events that mimic the imbalance in transverse momentum due to limitations in the STAR detector coverage in the region $\eta < -1$;
the background contributions from $W \rightarrow \tau + \nu$ and $Z \rightarrow ee$ decays are determined from MC simulation.
A good agreement is found between data and the combined MC contributions.
Finally, an additional restriction, $E_\mathrm{T} > 25 \unit{GeV}$, is applied in order to isolate the EW $e^\pm$ signal from the background.
On the other hand, the $Z$-boson kinematics can be reconstructed by requiring events to have two electrons with opposite charges.

The selected $W$ and $Z$ candidates are used to extract the $W$ and $Z$ fiducial cross sections as follows:

\begin{equation}
\sigma_{W^\pm,Z}^\mathrm{fid} = \frac{N_{W^\pm,Z}^O - N_{W^\pm,Z}^B}{\mathcal{L} \cdot \epsilon_{W^\pm,Z}},
\end{equation}

\noindent
where $N^O$ is the number of observed boson candidates, $N^B$ is the number of background events estimated from data and MC, $\mathcal{L}$ is the integrated luminosity, $\epsilon$ is the detection efficiency, and the subscript $W^\pm,Z$ refers to the respective boson candidate.

\begin{figure}[h]
\centering
\includegraphics[width=0.7\textwidth]{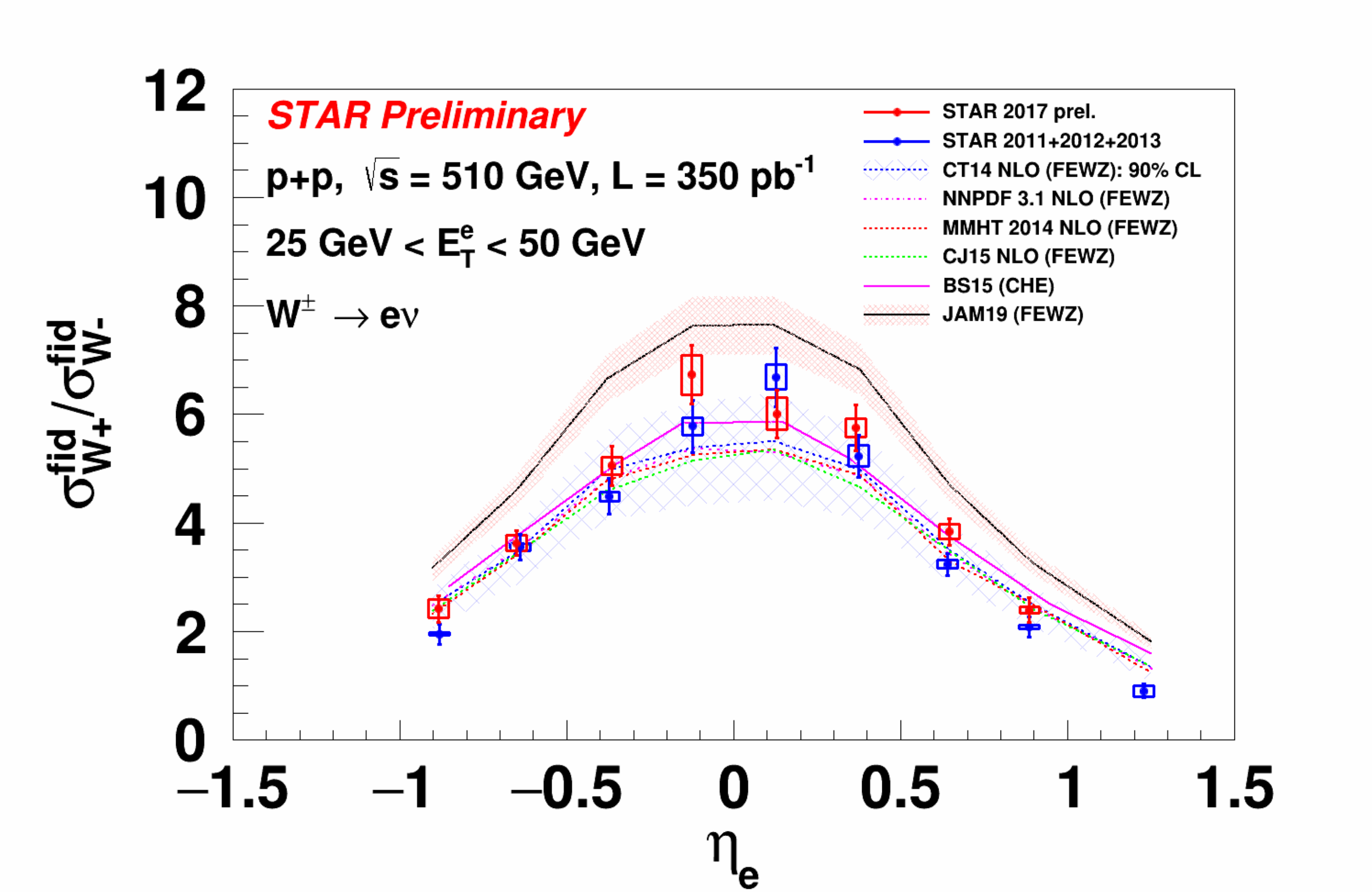}
\caption{
STAR $W$ cross-section ratios
from the 2011+2012+2013 data set and the 2017 data set plotted as a function of lepton pseudorapidity.
}
\label{fig:RW}
\end{figure}

Shown in Fig.~\ref{fig:RW} is a comparison of $W$ cross-section ratios from the 2011+2012+2013 data set and the 2017 data set as a function of lepton pseudorapidity.
The vertical bars represent the statistical uncertainties, while the boxes represent systematic uncertainties.
The colored band and curves correspond to theoretical calculations based on different PDF sets~\cite{Lai:2010vv,Bourrely:2001du} and frameworks~\cite{Gavin:2012sy,deFlorian:2010aa}.
The endcap measurement in the region $\eta_e > 1$ for the 2017 data set is still under analysis.
Only the 2011+2012+2013 result is shown.

\begin{figure}[h]
\centering
\includegraphics[width=0.45\textwidth]{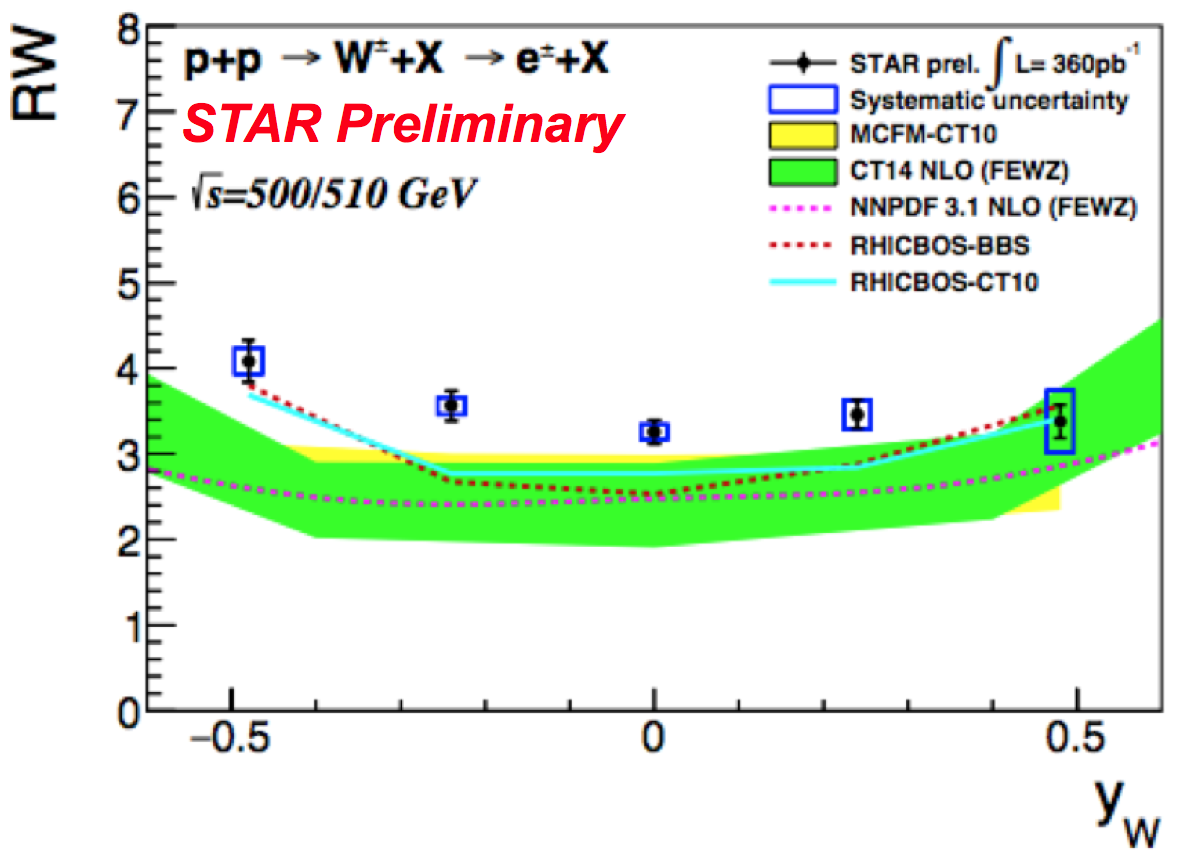}
\includegraphics[width=0.37\textwidth]{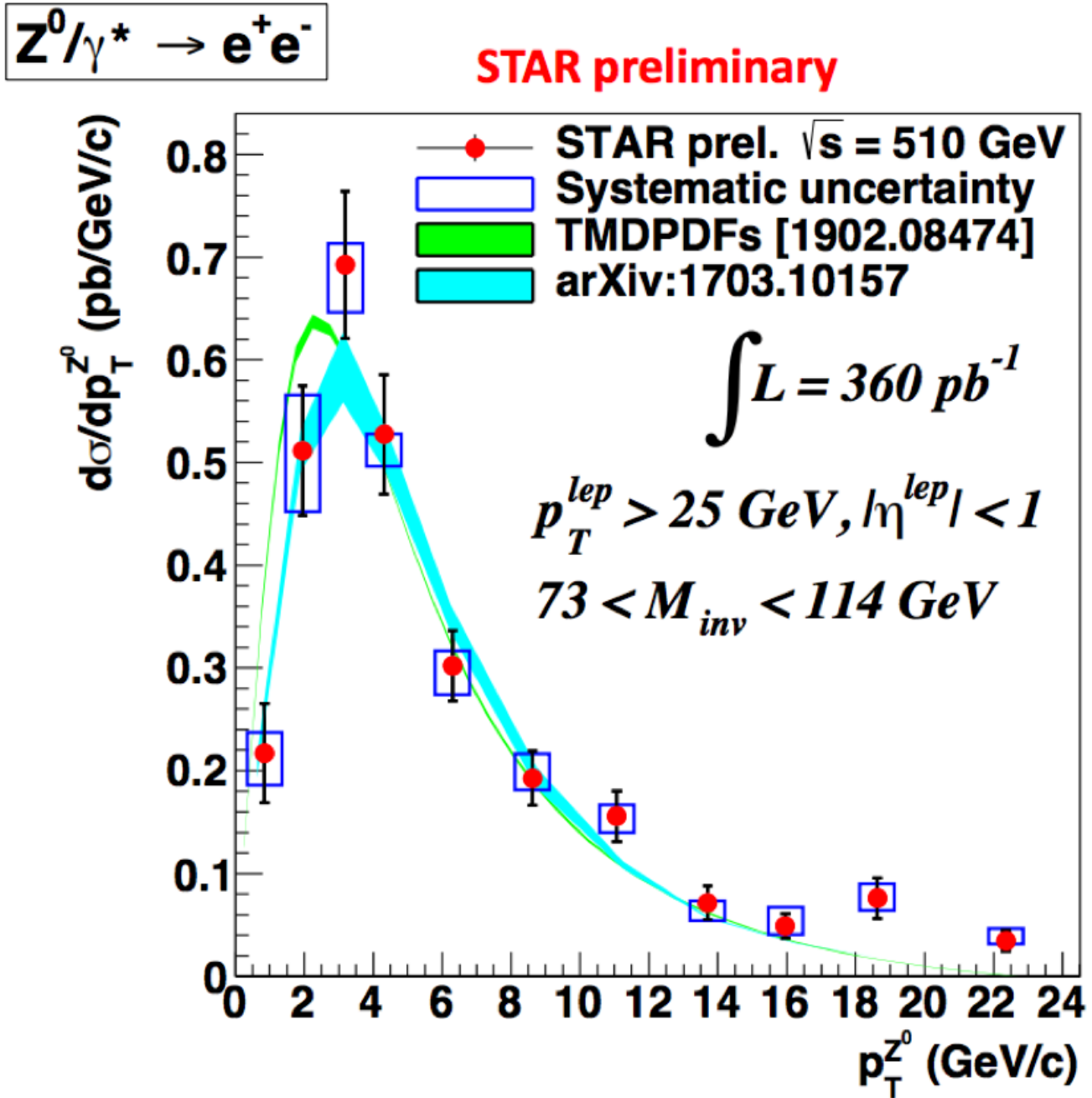}
\caption{
STAR $W$ cross-section ratios obtained with MC reconstruction method using the 2011+2012+2013 data set as a function of $W$ rapidity (left).
Differential cross section $d\sigma/dp_T^{Z^0}$ as a function of $Z^0$ transverse momentum $p_T^{Z^0}$ (right).
}
\label{fig:RW2}
\end{figure}

The $W$-boson kinematics can also be reconstructed at STAR, as described in~\cite{STAR:2015vmv}.
The neutrino momentum, $\vec{P}_\nu$, is determined from the imbalance in $p_\mathrm{T}$ with the help of the MC simulation.
The $R_W (\equiv \sigma_{W+}/\sigma_{W-})$ cross-section ratio obtained with the 2011+2012+2013 data set is illustrated in the left panel of Fig.~\ref{fig:RW2} as a function of reconstructed boson rapidity, $y_W$.

In addition, the differential cross section $d\sigma/dp_\mathrm{T}^{Z^0}$ has been measured with the 2011+2012+2013 data and the preliminary result is shown in the right panel of Fig.~\ref{fig:RW2}.
The result is compared to different TMDPDFs~\cite{Bertone:2019nxa,Bacchetta:2017gcc}.

\section{Conclusion}
\label{sec:conclusion}

STAR has measured the $W^+/W^-$ cross-section ratio in $pp$ collisions at $\sqrt{s} = 500 \unit{GeV}$ and $510 \unit{GeV}$. These measurements provide high $Q^2$ data that are sensitive to the $\bar{d}/\bar{u}$ ratio in the kinematic range of about $0.06 < x < 0.4$, which will help constrain the sea quark PDFs and complement the E866 and SeaQuest measurements.
Furthermore, the STAR results are complementary to the LHC $W$ and $Z$ production measurements by providing cross-section measurements at lower $\sqrt{s}$ and larger $x$.
Preliminary results for $W$ and $Z$ production based on the 2011+2012+2013 ($\mathcal{L} \sim 350 \unit{pb^{-1}}$) and 2017 ($\mathcal{L} \sim 350 \unit{pb^{-1}}$) data sets are presented, and compared to model calculations.

\section*{Acknowledgements}

We thank the RHIC Operations Group and RCF at BNL. This work is supported by U.S. DOE Office of Science and DOE-310385.



\bibliography{myref.bib}

\begin{thebibliography}{10}
\providecommand{\url}[1]{\texttt{#1}}
\providecommand{\urlprefix}{URL }
\expandafter\ifx\csname urlstyle\endcsname\relax
  \providecommand{\doi}[1]{doi:\discretionary{}{}{}#1}\else
  \providecommand{\doi}{doi:\discretionary{}{}{}\begingroup
  \urlstyle{rm}\Url}\fi
\providecommand{\eprint}[2][]{\url{#2}}

\bibitem{Gao:2013xoa}
J.~Gao, M.~Guzzi, J.~Huston, H.-L. Lai, Z.~Li, P.~Nadolsky, J.~Pumplin,
  D.~Stump and C.~P. Yuan,
\newblock \emph{{CT10 next-to-next-to-leading order global analysis of QCD}},
\newblock Phys. Rev. D \textbf{89}(3), 033009 (2014),
\newblock \doi{10.1103/PhysRevD.89.033009}.

\bibitem{Szydlowski:2008by}
M.~Szyd\l{}owski, A.~Krawiec, A.~Kurek and M.~Kamionka,
\newblock \emph{{AIC, BIC, Bayesian evidence against the interacting dark
  energy model}},
\newblock Eur. Phys. J. C \textbf{75}(99), 5 (2015),
\newblock \doi{10.1140/epjc/s10052-014-3236-1}.

\bibitem{Bourrely:2015kla}
C.~Bourrely and J.~Soffer,
\newblock \emph{{New developments in the statistical approach of parton
  distributions: tests and predictions up to LHC energies}},
\newblock Nucl. Phys. A \textbf{941}, 307 (2015),
\newblock \doi{10.1016/j.nuclphysa.2015.06.018}.

\bibitem{NuSea:2001idv}
R.~S. Towell \emph{et~al.},
\newblock \emph{{Improved measurement of the anti-d / anti-u asymmetry in the
  nucleon sea}},
\newblock Phys. Rev. D \textbf{64}, 052002 (2001),
\newblock \doi{10.1103/PhysRevD.64.052002}.

\bibitem{SeaQuest:2021zxb}
J.~Dove \emph{et~al.},
\newblock \emph{{The asymmetry of antimatter in the proton}},
\newblock Nature \textbf{590}(7847), 561 (2021),
\newblock \doi{10.1038/s41586-021-03282-z}.

\bibitem{Soffer:2014upa}
J.~Soffer, C.~Bourrely and F.~Buccella,
\newblock \emph{{Statistical description of the flavor structure of the nucleon
  sea}},
\newblock In \emph{{15th Workshop on High Energy Spin Physics}} (2014),
  \eprint{1402.0514}.

\bibitem{STAR:2002eio}
K.~H. Ackermann \emph{et~al.},
\newblock \emph{{STAR detector overview}},
\newblock Nucl. Instrum. Meth. A \textbf{499}, 624 (2003),
\newblock \doi{10.1016/S0168-9002(02)01960-5}.

\bibitem{Anderson:2003ur}
M.~Anderson \emph{et~al.},
\newblock \emph{{The Star time projection chamber: A Unique tool for studying
  high multiplicity events at RHIC}},
\newblock Nucl. Instrum. Meth. A \textbf{499}, 659 (2003),
\newblock \doi{10.1016/S0168-9002(02)01964-2}.

\bibitem{STAR:2002ymp}
M.~Beddo \emph{et~al.},
\newblock \emph{{The STAR barrel electromagnetic calorimeter}},
\newblock Nucl. Instrum. Meth. A \textbf{499}, 725 (2003),
\newblock \doi{10.1016/S0168-9002(02)01970-8}.

\bibitem{STAR:2002eml}
C.~E. Allgower \emph{et~al.},
\newblock \emph{{The STAR endcap electromagnetic calorimeter}},
\newblock Nucl. Instrum. Meth. A \textbf{499}, 740 (2003),
\newblock \doi{10.1016/S0168-9002(02)01971-X}.

\bibitem{STAR:2011aa}
L.~Adamczyk \emph{et~al.},
\newblock \emph{{Measurement of the $W \to e \nu$ and $Z/\gamma^* \to e^+e^-$
  Production Cross Sections at Mid-rapidity in Proton-Proton Collisions at
  $\sqrt{s}$ = 500 GeV}},
\newblock Phys. Rev. D \textbf{85}, 092010 (2012),
\newblock \doi{10.1103/PhysRevD.85.092010}.

\bibitem{STAR:2020vuq}
J.~Adam \emph{et~al.},
\newblock \emph{{Measurements of $W$ and $Z/\gamma^*$ cross sections and their
  ratios in p+p collisions at RHIC}},
\newblock Phys. Rev. D \textbf{103}(1), 012001 (2021),
\newblock \doi{10.1103/PhysRevD.103.012001}.

\bibitem{Sjostrand:2006za}
T.~Sjostrand, S.~Mrenna and P.~Z. Skands,
\newblock \emph{{PYTHIA 6.4 Physics and Manual}},
\newblock JHEP \textbf{05}, 026 (2006),
\newblock \doi{10.1088/1126-6708/2006/05/026}.

\bibitem{GEANT4:2002zbu}
S.~Agostinelli \emph{et~al.},
\newblock \emph{{GEANT4--a simulation toolkit}},
\newblock Nucl. Instrum. Meth. A \textbf{506}, 250 (2003),
\newblock \doi{10.1016/S0168-9002(03)01368-8}.

\bibitem{Lai:2010vv}
H.-L. Lai, M.~Guzzi, J.~Huston, Z.~Li, P.~M. Nadolsky, J.~Pumplin and C.~P.
  Yuan,
\newblock \emph{{New parton distributions for collider physics}},
\newblock Phys. Rev. D \textbf{82}, 074024 (2010),
\newblock \doi{10.1103/PhysRevD.82.074024}.

\bibitem{Bourrely:2001du}
C.~Bourrely, J.~Soffer and F.~Buccella,
\newblock \emph{{A Statistical approach for polarized parton distributions}},
\newblock Eur. Phys. J. C \textbf{23}, 487 (2002),
\newblock \doi{10.1007/s100520100855}.

\bibitem{Gavin:2012sy}
R.~Gavin, Y.~Li, F.~Petriello and S.~Quackenbush,
\newblock \emph{{W Physics at the LHC with FEWZ 2.1}},
\newblock Comput. Phys. Commun. \textbf{184}, 208 (2013),
\newblock \doi{10.1016/j.cpc.2012.09.005}.

\bibitem{deFlorian:2010aa}
D.~de~Florian and W.~Vogelsang,
\newblock \emph{{Helicity Parton Distributions from Spin Asymmetries in W-Boson
  Production at RHIC}},
\newblock Phys. Rev. D \textbf{81}, 094020 (2010),
\newblock \doi{10.1103/PhysRevD.81.094020}.

\bibitem{STAR:2015vmv}
L.~Adamczyk \emph{et~al.},
\newblock \emph{{Measurement of the transverse single-spin asymmetry in
  $p^\uparrow+p \to W^{\pm}/Z^0$ at RHIC}},
\newblock Phys. Rev. Lett. \textbf{116}(13), 132301 (2016),
\newblock \doi{10.1103/PhysRevLett.116.132301}.

\bibitem{Bertone:2019nxa}
V.~Bertone, I.~Scimemi and A.~Vladimirov,
\newblock \emph{{Extraction of unpolarized quark transverse momentum dependent
  parton distributions from Drell-Yan/Z-boson production}},
\newblock JHEP \textbf{06}, 028 (2019),
\newblock \doi{10.1007/JHEP06(2019)028}.

\bibitem{Bacchetta:2017gcc}
A.~Bacchetta, F.~Delcarro, C.~Pisano, M.~Radici and A.~Signori,
\newblock \emph{{Extraction of partonic transverse momentum distributions from
  semi-inclusive deep-inelastic scattering, Drell-Yan and Z-boson production}},
\newblock JHEP \textbf{06}, 081 (2017),
\newblock \doi{10.1007/JHEP06(2017)081},
\newblock [Erratum: JHEP 06, 051 (2019)].

\end{thebibliography}

\nolinenumbers

\end{document}